\documentclass[conference]{IEEEtran}

\usepackage[T1]{fontenc}
\usepackage{lmodern}
\usepackage{textcomp}
\usepackage{latexsym}
\usepackage[fleqn]{amsmath}
\usepackage{bm} 
\usepackage{relsize}
\usepackage{amsmath}
\usepackage{amsfonts}
\usepackage{empheq}
\usepackage{cases}

\usepackage[pdftex]{graphicx,xcolor}
\usepackage{url}
\usepackage{setspace}
\usepackage{caption}
\usepackage{theorem}
\usepackage[labelformat=simple]{subcaption}

\usepackage{latexsym}
\usepackage{amssymb}
\usepackage{algorithm}
\usepackage{algorithmic}
\usepackage{epic,eepic}

\renewcommand{\baselinestretch}{1}

\newtheorem{remark}{Remark}

\ifCLASSINFOpdf
\else
\fi

\begin{document}
\title{Efficient Scheduling of Serial Iterative Decoding for Zigzag Decodable Fountain Codes}

\author{\IEEEauthorblockN{Yoshihiro Murayama and Takayuki Nozaki}
\IEEEauthorblockA{
Dept.\ of Informatics, Yamaguchi University, JAPAN \\
Email: {\tt \{i033vb,tnozaki\}@yamaguchi-u.ac.jp}}
}

\maketitle

\begin{abstract}
Fountain codes are erasure correcting codes realizing reliable communication systems for the multicast on the Internet.
The zigzag decodable fountain (ZDF) code is one of generalization of the Raptor code, i.e, applying shift operation to generate the output packets.
The ZDF code is decoded by a two-stage iterative decoding algorithm, which combines the packet-wise peeling algorithm and the bit-wise peeling algorithm.
By the bit-wise peeling algorithm and shift operation, ZDF codes outperform Raptor codes under iterative decoding in terms of decoding erasure rate and overhead.
However, the bit-wise peeling algorithm spends long decoding time.
This paper proposes a fast bit-wise decoding algorithm for the ZDF codes.
Simulation results show that the proposed algorithm drastically reduces the decoding time compared with the existing algorithm.
\end{abstract}

\IEEEpeerreviewmaketitle

\section{Introduction}
On the Internet, each message is split into several packets.
We regard the packets which are not correctly received as erased.
Hence, the Internet is modeled as a packet erasure channel.
The sender cannot retransmit packets in the case of user datagram protocol (UDP). 

Fountain codes \cite{FountainCode} are erasure correcting codes which realizes reliable communication via UDP, in particular, multicasting.
We assume that the original message is split into $k$ source packets.
In the fountain coding system, the sender generates infinite output packets from the $k$ source packets.
Each receiver decodes the original message from $k(1+\alpha)$ received packets, where $\alpha$ is referred to as {\it packet overhead}.
Hence, in the fountain coding system, the receiver need not request the retransmission.

Raptor code \cite{RaptorCode} is a fountain code which achieves arbitrarily small $\alpha$ as $k\to \infty$ with a linear time encoding and decoding algorithm.
Encoding of Raptor code is divided into two stages.
At the first stage, the encoder generates {\it precoded packets} from the source packets by using a precode, which is a high rate erasure correcting code, e.g, low-density parity-check (LDPC) code.
At the second stage, the encoder generates output packets from precoded packets by using LT code \cite{LTCode}.
More precisely, each output packet is generated from the bit-wise exclusive OR (XOR) of randomly chosen precoded packets. 
Decoding of Raptor codes is in a similar way to LDPC code over the binary erasure channels. 
In other words, the decoder constructs the factor graph from the received packets and the parity check matrix of the precode and recovers the precoded packets by using the peeling algorithm (PA) \cite{PA}.

Zigzag decodable fountain (ZDF) code \cite{ZDF} is a generalization of Raptor code.
Similarly to Raptor code, encoding of ZDF code is divided into two stages.
At the first stage, the encoder generates precoded packets from the source packets by using a precode.
At the second stage, the encoder generates output packets from the precoded packets in the following way; 
Encoder randomly chooses precoded packets and those shift amounts, executes the bit-level shift to the chosen precoded packets, and perform the bit-wise XOR to the shifted precoded packets.

A decoding algorithm for the ZDF codes is also two-stage algorithm.
Similarly to Raptor code, a factor graph for the ZDF code is constructed before stating the decoding algorithm.
At the first stage, a packet-wise PA works on the factor graph and recovers the precoded packets in packet-wise.
Unless the packet-wise PA succeeds, the remaining precoded packets are decoded by the bit-wise PA, which is the PA over the bit-wise representation of the factor graph. 

As shown in \cite{ZDF}, ZDF codes outperform Raptor codes in terms of packet overhead.
However, the decoding algorithm for ZDF codes requires large decoding time.
The purpose of this research is to propose a fast decoding algorithm for ZDF codes.
As a related work, the work in \cite{ZDF_TEP} slightly reduces the number of decoding iterations.

In this paper, we propose a fast decoding algorithm for ZDF codes by reducing the number of decoding processes in the bit-wise PA.
By a numerical example shown in Section \ref{sec:con_ori}, we ascertain that only particular edges contribute to recovering the bits of precoded packets.
The main idea of this work is to execute the decoding processes only to such edges.
Moreover, the algorithm makes a list which records the order of edges which contribute the decoding and execute the decoding processes as in the order of this list.
As a result, we significantly reduce the decoding time compared with the existing algorithm.

The rest of paper is organized as follows.
Section \ref{sec:pre} briefly reviews ZDF codes and those existing decoding algorithm.
Section \ref{sec:new} gives a numerical example which shows that the only particular edges contribute the decoding, and proposes a fast decoding algorithm for the ZDF codes via scheduling.
The simulation results in Section \ref{sec:kekka} show that the proposed algorithm reduces the number of decoding processes and decoding time compared with exist one.
Section \ref{sec:5} concludes the paper.

\section{Preliminaries}\label{sec:pre}
This section gives some notation and introduces the encoding and decoding algorithm of the ZDF codes.
Section \ref{sec:ZDF} explains the encoding of the ZDF codes.
Section \ref{sec:ZDF_fac} gives factor graph representations of the ZDF codes.
Section \ref{sec:ZDF_Dec} explains the original decoding algorithm \cite{ZDF} of the ZDF codes.

\subsection{Encoding of the ZDF Codes \label{sec:ZDF} \cite{ZDF}}
A polynomial representation of the packets $\bm{a}=(a_1,a_2,\dots,a_\ell)$ is defined as $a(z)=\sum^\ell_{j=1}a_jz^j$.

The ZDF codes is defined by precode $\mathcal{C}$, the distribution for the inner code $\Omega (x)=\sum_i\Omega_ix^i$, and shift distribution $\Delta(x)=\sum^D_{i=0}\Delta_ix^i$.
Here, $\Delta_i$ represents the probability that the shift amount is $i$.

Similarly to the Raptor codes, the ZDF codes
generates the precoded packets $\bm{b}_1,\bm{b}_2,\dots,\bm{b}_n$
from the source packets $\bm{a}_1,\bm{a}_2,\dots,\bm{a}_k$ by the precode $\mathcal{C}$ at the first stage.
At the second stage, the ZDF codes generates the infinite
output packets as the following procedure for $t = 1,2,\dots$.
\begin{enumerate}
\item Choose a degree $d$ of the $t$-th output packet according to the degree distribution $\Omega(x)$. In other words, choose $d$ with probability $\Omega_d$.
\item Choose $d$-tuple of shift amounts $(\tilde{\delta}_{t,1},\tilde{\delta}_{t,2},\dots,\tilde{\delta}_{t,d})\in[0,D]^d := \{0,1,\dots,D\}^d$ in independent of each other according to shift distribution $\Delta(x)$, where $[a, b]$ denote the set of integers between $a$ and $b$. Define  $\tilde{\delta}_{t,\min} := \min_i\tilde{\delta}_{t,i}$ and calculate $\delta_{t,i} := \tilde{\delta}_{t,i} - \tilde{\delta}_{t,\min}$.
\item Choose $d$ distinct precoded packets uniformly. Let $(j_1,\allowbreak j_2,\allowbreak \dots,\allowbreak j_d)$ denote the $d$-tuple of indexes of the chosen precoded packets. Then the polynomial representation for the $t$-th output packet is given as
  \begin{eqnarray*}
    x_t(z)=\sum^d_{i=1}z^{\delta_{t,i}}b_{j_i}(z).
  \end{eqnarray*}
\end{enumerate}

\subsection{Factor graph generated by the receiver \label{sec:ZDF_fac}}
Let $\bm{y}_1,\bm{y}_2,\dots,\bm{y}_{k'}$ be $k'$ received packets for a receiver, where $k' := k(1+\alpha)$.
Similarly to the Raptor codes, each receiver constructs a factor graph from the precode $\mathcal{C}$ and the received packets.
The generated factor graphs depend on receivers since the $k'$ received packets depend on receivers.

The factor graphs for the ZDF codes is composed of labeled edges and the four kinds of nodes: $n$ variable nodes representing precoding packets $\mathsf{V}_{\mathrm{p}} := \{\mathsf{v}_1,\mathsf{v}_2,\dots,\mathsf{v}_n\}$, $m := n-k$ check nodes on the precode code $\mathsf{F}_{\mathrm{p}} := \{\mathsf{f}_1,\dots,\mathsf{f}_m\}$, $k'$ variable nodes representing received packets $\mathsf{V}_{\mathrm{r}} := \{\mathsf{v}_{1}', \allowbreak \mathsf{v}_{2}', \allowbreak \dots, \allowbreak \mathsf{v}_{k'}'\}$, and $k'$ factor nodes on the inner code $\mathsf{F}_{\mathrm{i}} := \{\mathsf{f}_{m+1},\dots,\mathsf{f}_{m+k'}\}$.

The edge connection between $\mathsf{F}_{\mathrm{p}}$ and $\mathsf{V}_{\mathrm{p}}$ is decided from the precode $\mathcal{C}$.
More precisely, $\mathsf{f}_{i}$ and $\mathsf{v}_{j}$ are connected to an edge labeled by 1 if and only if the $(i, j)$-th entry of $\mathcal{C}$ is equal to 1.
The edge connection between $\mathsf{F}_{\mathrm{i}}$ and $\mathsf{V}_{\mathrm{p}}$ is decided from the header of the $i$-th received packet.
If the header of the $i$-th received packet represents $(\delta_1,\delta_2,\dots,\delta_d)$ and $(j_1,j_2,\dots,j_d)$, an edge labeled by $z^{\delta_{t,i}}$ connects $\mathsf{f}_{m+t_i}$ and $\mathsf{v}_{j_i}$ for $i \in [1,d]$.
We denote the label on the edges connecting to $\mathsf{f}_{m+i}$ and $\mathsf{v}_{j}$, by $z^{\delta_{i,j}}$. 
For $i \in [1,k']$, an edge connects $\mathsf{f}_{m+i}$ and $\mathsf{v}_{i}'$.
Denote the set of indexes of the variable nodes adjacent to the $i$-th factor node, by $\mathcal{N}_{f}(j)$.

\subsection{Decoding algorithm for the ZDF codes\label{sec:ZDF_Dec}}

The decoding algorithm of the ZDF codes is a two-stage algorithm.
At the first stage, packet-wise PA works on the factor graph of the ZDF codes.
The details of packet-wise PA is given in \cite{ZDF}.
Unless the packet-wise PA successes, bit-wise PA works on its residual graph.
In this section, we explain the original bit-wise PA \cite{ZDFalgo}.

In the decoding of the ZDF codes, the $i$-th factor node has memory of length $\ell+D$, denoted by $\bm{s}_i \in \{0,1\}^{\ell+D}$.
Let $\mathcal{E}$ be the set of indexes of variable nodes which are not recovered by the packet-wise PA.
The residual graph is the subgraph composed of the variable nodes in $\mathcal{E}$ and those connecting edges.

Now, we explain factor node processing of bit-wise PA.
Let $(w_{t,1},\dots,w_{t,\ell})$ be vector representation of packet $\bm{w}_t$.
Denote the mapping of the factor node processing, by $\Phi '$.
The mapping $\Phi'$ updates the memory $\bm{b}_j$ of the $j$-th variable node by the $i$-th factor node as follows
\begin{align*}
&\Phi ' (i,j,\{\bm{b}_t\}_{t \in \mathcal{N}_{f}(i)})\\
=&\Psi(b_j,\mathcal{S}^{+}(\delta_{i,j},\Phi(\bm{s}_i,\{\mathcal{S}(\delta_{i,t},\bm{b}_t)\}_{t\in \mathcal{N}_{f}(i)\setminus\{j\}}))),
\end{align*}
where the mapping $\mathcal{S}:[0,D]\times\{0,1,*\}^\ell\to \{0,1,*\}^{\ell+D}$ is
\[
\mathcal{S}(k,\bm{w}_t)
=
(\overset{k}{\overbrace{0,\dots,0}},w_{t,1},w_{t,2},\dots,w_{t,l},\overset{D-k}{\overbrace{0,\dots,0}}),
\]
the mapping $\Phi:\{0,1,*\}^{(\ell+D)\times d_v}\to\{0,1,*\}^{\ell+D}$ is
\begin{align*}
&\tilde{\bm{w}}_0=\Phi(\tilde{\bm{w}}_1,\dots,\tilde{\bm{w}}_{d_v}),
k\in[1,\ell+D],\\
&\tilde{w}_{0,k} =
\begin{cases}
  \sum_{i=1}^{d_v}\tilde{w}_{i,k}, & \text{if} \hspace{2mm}\tilde{w}_{1,k},\dots,\tilde{w}_{d_v,k}\in\{0,1\}, \\
  *, & \text{otherwise},
\end{cases}
\end{align*}
the mapping $\mathcal{S}^+(k,\cdot) : \{0,1,* \}^{\ell+D} \to \{0, 1, * \}^\ell$ is
\[
\mathcal{S}^+(k,\tilde{\bm{w}}_t)
= 
(\tilde{w}_{t,k+1},\tilde{w}_{t,k+2},\dots,\tilde{w}_{t,k+\ell}),
\]
and the mapping $\Psi:\{0,1,*\}^{\ell \times 2}\to\{0,1,*\}^{\ell}$ is
\begin{align*}
  &\tilde{\bm{w}}_0=\Psi(\tilde{\bm{w}}_1,\tilde{\bm{w}}_{2}),k\in[1,\ell+D]\\
&\tilde{w}_{0,k} =
\begin{cases}
  \tilde{w}_{1,k}, & \text{if} \hspace{2mm}\tilde{w}_{1,k}\in\{0,1\}, \\
  \tilde{w}_{2,k}, & \text{otherwise}.
\end{cases}
\end{align*}

Bit-wise decoding is done in the procedure of Algorithm \ref{algc:2}.
\begin{algorithm}[tb]
  \begin{footnotesize}
    \caption{Bit-wise decoding (existing algorithm) \label{algc:2}}
    \begin{algorithmic}[1]
      \REQUIRE Residual graph $\mathtt{G}$, values of memories $\bm{s}_i$ $(i \in [1,m+k'])$, and precoded packets $\bm{b}_1,\dots,\bm{b}_n$
      \ENSURE precoded packets $\bm{b}_1,\dots,\bm{b}_n$
      \STATE $\tau \gets 1$,
      \STATE $\forall j \in [1,n]~ \bm{b}_j^{(0)} \gets \bm{b}_j$
      \FOR{$i\in[1,m+k^{\prime}]$, $j \in \mathcal{N}_f(i)$}\label{stp:rec2}
      \STATE $\bm{b}_j \gets \Phi ' (i,j,\{\bm{b}_t\}_{t \in \mathcal{N}_{f}(i)})$\label{stp:PPP}
      \ENDFOR 
      \STATE $\forall j \in [1,n]~ \bm{b}_j^{(\tau)} \gets \bm{b}_j$
      \IF{$\forall j \in [1,n]~ \bm{b}_j^{(\tau)} \neq \bm{b}_j^{(\tau-1)}$}
      \STATE $\tau \gets \tau +1$ and go to Step \ref{stp:rec2}
      \ENDIF
    \end{algorithmic}
  \end{footnotesize}
\end{algorithm}
If Algorithm \ref{algc:2} outputs $b_j(z) \in \{0,1\}^\ell$ for all $j \in [1,n]$, the decoding succeeds.
Otherwise, decoding fails.
\section{Observation of the bit-wise PA and Proposed Decoding Algorithm}\label{sec:new}
We refer to the process of Step \ref{stp:PPP} in Algorithm \ref{algc:2} as {\it decoding process}.
We refer to the edges which recover the bits of precodes as {\it updating edges}.
In other words, the updating edges contribute the decoding at a decoding round.
In the original bit-wise PA, since all the factor nodes are processed, the number of decoding processes per iteration is equal to the total number of edges of the factor graph generated by the receiver.
Hence, the decoding process is performed on many edges which do not contribute to decoding.

In this section, we observe the original bit-wise decoding.
As a result, we ascertain that the only particular edges contribute the decoding. 

Roughly speaking, the proposed bit-wise decoding algorithm generates a set of edges used for updating variable nodes and records the edges to a list in the proper order from the set of edges.
After that, the decoding process is performed only on edges in the list.

Section \ref{sec:con_ori} shows that the only particular edges contribute the decoding by the evaluation of original bit-wise PA.
Section \ref{sec:new_dec} gives proposed decoding algorithm which reduces the decoding process per iteration.

\subsection{Decoding Process of the original bit-wise PA \label{sec:con_ori}}
In this section, we use the shift distribution $\Delta(x) = \frac{1}{2}+\frac{1}{2}x$.
As a precode, we employ (3,30)-regular LDPC codes.
The degree distribution for the inner code is $\Omega(x) 
 = 
0.007969x + 0.493570x^2 + 0.166220x^3 + 0.072646x^4+0.082558x^5
+ 0.056058x^8+0.037229x^9+0.055590x^{19}+0.025023x^{65}+0.003135x^{66}$.
An edge is {\it activate} if the edge has become an updating edge.

\begin{figure}[t]
  \centering
    \includegraphics[width=.835\linewidth]{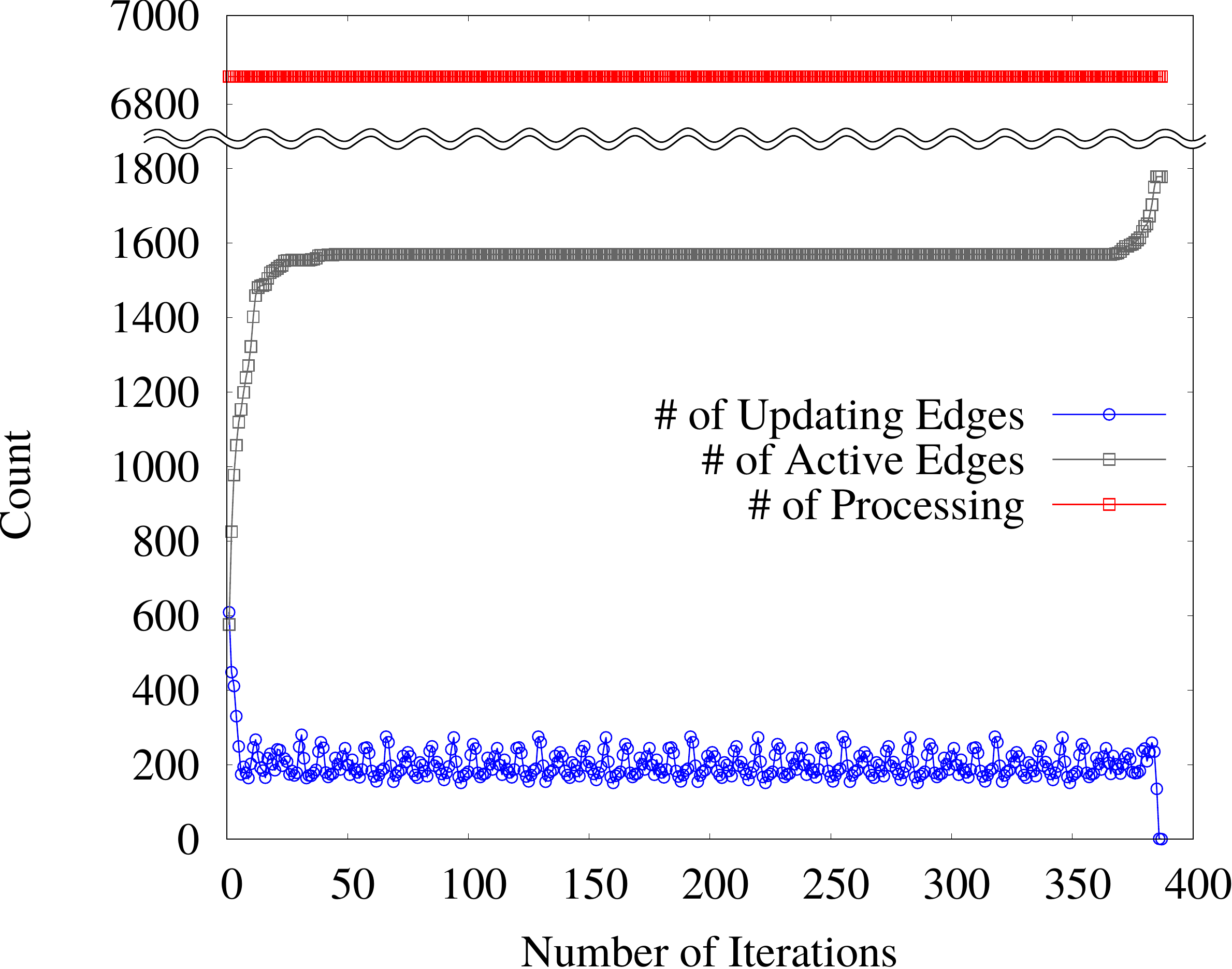}
  \caption{Number of decoding processes, updating edges, and activated edges per iteration (existing algorithm) \label{fig:num_proc}}
\end{figure}

Figure \ref{fig:num_proc} displays the number of decoding processes, the number of updating edges, and the number of active edges under the original bit-wise PA with $\ell=100$ and $\alpha=0.10$.
The horizontal axis of Fig.~\ref{fig:num_proc} represents the number of iterations.
The symbol \# in Fig.~\ref{fig:num_proc} stands ``the number''.
From the curve of active edges in Fig.~\ref{fig:num_proc}, we see that the only particular edges contribute the decoding.

\subsection{Proposed Algorithm \label{sec:new_dec}}
Let $\mathcal{L}^{\mathsf{A}}$ (resp.\ $\mathcal{L}^{\mathsf{B}}$) be the set (resp.\ list) of edges which contribute the decoding by the original bit-wise PA.
In other words, $\mathcal{L}^{\mathsf{A}}$ represents the set of active edges.
Proposed decoding algorithm is divided into three stages.
At the first stage, the decoder executes decoding process for all the edges and makes $\mathcal{L}^{\mathsf{A}}$ by adding the edges which contribute the decoding of precoded packets.
At the second stage, the decoder executes decoding process for the edges in $\mathcal{L}^{\mathsf{A}}$ and makes $\mathcal{L}^{\mathsf{B}}$ by recording the order of edges which contribute the decoding.
At the final stage, the decoding process is performed edges as in the order of the list $\mathcal{L}^{\mathsf{B}}$.
Unless edges contribute the decoding, the edges is deleted from the list $\mathcal{L}^{\mathsf{B}}$.
\begin{remark} \upshape \label{rem:0}
At the second stage, decoder makes only one list $\mathcal{L}^{\mathsf{B}}$.
  The $j$-th element $\mathcal{L}^{\mathsf{B}}_j$ of list $\mathcal{L}^{\mathsf{B}}$ stores the $j$-th contributing edge in the whole second stage.
  Hence, there is a possibility that an edge is recorded in $\mathcal{L}^{\mathsf{B}}$ several times.
\end{remark}

Next, we explain the parameters used in the proposed algorithm. 
Let $l_{\mathsf{A}}$ and $l_{\mathsf{B}}$ represent the size of $\mathcal{L}^{\mathsf{A}}$ and the length of $\mathcal{L}^{\mathsf{B}}$, respectively.
The number of iterations in the second stage is denoted by $t_{\mathsf{B}}$.
We use time $t_{\mathsf{A}}$ and vector $\bm{V} = (V_1,V_2,\dots,V_{n})$ to reduce the time of make $\mathcal{L}^{\mathsf{A}}$.
The element $V_j$ of vector $\bm{V}$ indicates whether the $j$-th variable node connects to an edge in $\mathcal{L}^{\mathsf{A}}$.
More precisely, $V_j = 0$, if the  $j$-th variable node is already recovered or can be updated by an edge in $\mathcal{L}^{\mathsf{A}}$.
Otherwise, $V_j = 1$.
Function $T$ determines whether the variable node has been recovered, namely, 
\begin{equation*}
  T(\bm{b}_j) =
  \begin{cases}
    0, & \text{if} \hspace{2mm}\bm{b}_j\in\{0,1\}^{\ell}, \\
    1, & \text{otherwise}.
  \end{cases}
\end{equation*}
We denote the maximum number of decoding iterations at the first stage, by $t_{\mathsf{A}}$.
At the first stage, if there exists $j$ such that $V_j = 1$ until the $t_{\mathsf{A}}$-th iteration, then we expect that the $j$-th precoded packet will not be recovered.
Hence, in such case, the decoder halts and outputs decoding failure.

\begin{remark} \upshape \label{rem:1}
Time $t_{\mathsf{A}}$ is used time out of decoding.
Small $t_{\mathsf{A}}$ reduces average decoding time but slightly degrades the decoding performance.
Conversely, large $t_{\mathsf{A}}$ causes large average decoding time but does not degrade the decoding performance.
We confirm these in Section \ref{sec:hyo_er}.
\end{remark}

The details of the proposed algorithm are given in Algorithm \ref{algc:6}.
Step \ref{stp:LAcreate}-\ref{stp:end_LAcreate} gives the first stage of decoding.
Step \ref{stp:LAbranch1} decides decoding stop.
Step \ref{stp:LAbranch2} decides whether making of $\mathcal{L}^{\mathsf{A}}$ is sufficient.
Step \ref{stp:LBcreate}-\ref{stp:end_LBcreate} gives the second stage of decoding.
Step \ref{stp:LBuse}-\ref{stp:end_LBuse} gives the final stage of decoding.
At the final stage, the edges that do not contribute the decoding is deleted.
In Step \ref{stp:end_end}, if there exists an unrecovered precoded packet, then decoding restarts the first stage. 
Here, by this process, the decoding performance is improved.

\begin{algorithm}[t]
  \begin{footnotesize}
    \caption{Scheduled bit-wise decoding \label{algc:6}}
    \begin{algorithmic}[1]
      \REQUIRE Residual graph $\mathtt{G}$, values of memories $\bm{s}_i$ $(i \in [1,m+k'])$,
      precoded packets $\bm{b}_1,\dots,\bm{b}_n$, time $t_{\mathsf{A}}$, and time $t_{\mathsf{B}}$
      \ENSURE precoded packets $\bm{b}_1,\dots,\bm{b}_n$
      \setlength{\columnsep}{4pt}
      \STATE $\tau \gets 0$, $l_{\mathsf{A}} \gets 0$, $l_{\mathsf{B}} \gets 0$, $\forall j \in [1,n]~ \bm{b}_j^{(\tau)} \gets \bm{b}_j$
      \STATE $\tau_M \gets \tau + t_{\mathsf{A}}$, $\forall j~V_j \gets T(\bm{b}_j)$ \label{stp:start}
      \STATE $\tau \gets \tau + 1$\label{stp:LAcreate}
      \FOR{$i\in[1,m+k^{\prime}]$, $j \in \mathcal{N}_f(i)$} 
      \STATE $\bm{d} \gets \Phi(i,j,\{\bm{b}_t\}_{t \in \mathcal{N}_{f}(i)})$
      \IF{$\bm{d} \neq \bm{b}_j$}
      \STATE $V_j \gets 0$
      \IF{$ q \in [0,l_{\mathsf{A}}-1]~ \mathcal{L}^{\mathsf{A}}_{q} \neq (\mathsf{v}_j,\mathsf{f}_i)$}
      \STATE $\mathcal{L}^{\mathsf{A}}_{l_{\mathsf{A}}} \gets (\mathsf{v}_j,\mathsf{f}_i)$
      \STATE $l_{\mathsf{A}} \gets l_{\mathsf{A}}+1$
      \ENDIF
      \ENDIF
      \STATE $\bm{b}_j \gets \bm{d}$
      \ENDFOR 
      \STATE $\forall j \in [1,n]~ \bm{b}_j^{(\tau)} \gets \bm{b}_j$
      \IF{$\forall j\in[1,n]~ \bm{b}_j^{(\tau)} = \bm{b}_j^{(\tau-1)} $ or $\tau \geq \tau_M$}\label{stp:LAbranch1} 
      \STATE Decoding halts.
      \ELSIF{$\exists j~ V_j=1$}\label{stp:LAbranch2}
      \STATE Go to Step \ref{stp:LAcreate}.
      \ENDIF \label{stp:end_LAcreate}
      \STATE $\tau_M \gets \tau +t_{\mathsf{B}}$, $\forall j~V_j \gets T(\bm{b}_j)$
      \STATE $\tau \gets \tau + 1$\label{stp:LBcreate}
      \FOR{$q\in[0,l_{\mathsf{A}}-1]$}
      \STATE Set $j,i$ s.t.~$\mathcal{L}^{\mathsf{A}}_q = (\mathsf{v}_j,\mathsf{f}_i)$
      \STATE $\bm{d} \gets \Phi(i,j,\{\bm{b}_t\}_{t \in \mathcal{N}_{f}(i)})$
      \IF{$\bm{d} \neq \bm{b}_j$}
      \STATE $V_j \gets 0$, $\mathcal{L}^{\mathsf{B}}_{l_{\mathsf{B}}} \gets \mathcal{L}^{\mathsf{A}}_q$
      \STATE $l_{\mathsf{B}} \gets l_{\mathsf{B}}+1$
      \ENDIF
      \STATE $\bm{b}_j \gets \bm{d}$
      \ENDFOR
      \STATE $\forall j \in [1,n]~ \bm{b}_j^{(\tau)} \gets \bm{b}_j$
      \IF{$\forall j \in [1,n]~ \bm{b}_j^{(\tau)} = \bm{b}_j^{(\tau-1)}$ or ($\tau \geq \tau_M$ and $\exists j~V_j=1$)}
      \STATE  $l_{\mathsf{B}} \gets 0$ and go to Step \ref{stp:start}.
      \ELSIF{$\tau < \tau_M$}
      \STATE Go to Step \ref{stp:LBcreate}.
      \ENDIF \label{stp:end_LBcreate}
      \STATE $\tau \gets \tau +1$\label{stp:LBuse}
      \FOR{$q\in[0,l_{\mathsf{B}}-1]$}
      \STATE Set $j,i$ s.t.~$\mathcal{L}^{\mathsf{B}}_q = (\mathsf{v}_j,\mathsf{f}_i)$
      \STATE $\bm{d} \gets \Phi(i,j,\{\bm{b}_t\}_{t \in \mathcal{N}_{f}(i)})$
      \IF {$\bm{d} = \bm{b}_j$ }
      \STATE Delete $\mathcal{L}^{\mathsf{B}}_q$ from $\mathcal{L}^{\mathsf{B}}$.
      \ENDIF
      \STATE $\bm{b}_j \gets \bm{d}$
      \ENDFOR\label{stp:end3}
      \STATE $\forall j \in [1,n]~ \bm{b}_j^{(\tau)} \gets \bm{b}_j$
      \IF{$\exists j \in [1,n]~ \bm{b}_j^{(\tau)} \neq \bm{b}_j^{(\tau-1)}$}
      \STATE Go to Step \ref{stp:LBuse}\label{stp:end_LBuse}
      \ELSIF{$\exists j \in [1,n]~ \bm{b}_j \notin \{0,1\}^\ell$}
      \STATE Go to Step \ref{stp:start}\label{stp:end_end}
      \ENDIF
    \end{algorithmic}
  \end{footnotesize}
\end{algorithm}

If Algorithm \ref{algc:6} outputs $b_j(z) \in \{0,1\}^\ell$ for all $j \in [1,n]$, the decoding succeeds.
Otherwise, decoding fails.
\section{Simulation Results}\label{sec:kekka}
In this section, we evaluate the performance of the proposed algorithm.
Section \ref{sec:hyo_er} evaluates the decoding erasure rates.
Section \ref{sec:hyo_con} compares the number of decoding processes.
Section \ref{sec:hyo_time} gives the decoding time.
The parameters (i.e, $\mathcal{C},\Omega(x)$, and $\Delta(x)$) used in this section are same as Section \ref{sec:con_ori}.
The time $t_{\mathsf{A}},t_{\mathsf{B}}$ are given by $t_{\mathsf{A}} = 6/\alpha, t_{\mathsf{B}} = 20$.

\subsection{Decoding Erasure Rate \label{sec:hyo_er}}

\begin{figure}[t]
  \centering
  \includegraphics[width=.785\linewidth]{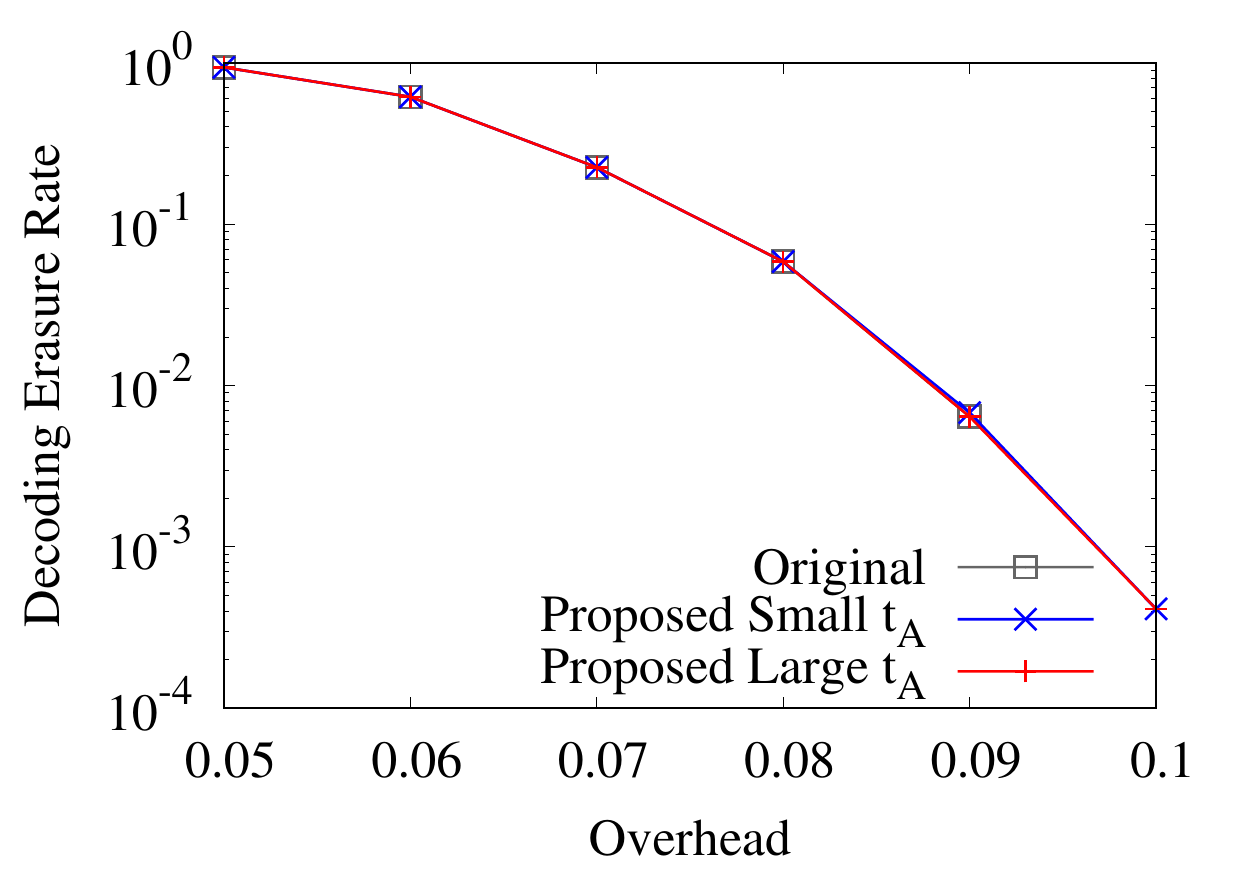}
  \caption{Comparison of decoding erasure rate of the proposed algorithm with existing one (Original) \label{fig:dec_er}}
\end{figure}
The decoding erasure rate (DER) is the fraction of the trials in which some bits in the precoded packets are not recovered.

Figure \ref{fig:dec_er} displays the DERs for the proposed algorithm and existing algorithm with $\ell=1000$.
The horizontal axis of Fig.~\ref{fig:dec_er} represents the packet overhead $\alpha$.
The curve labeled with ``Original'' in Fig.~\ref{fig:dec_er} gives the DER of existing algorithm, and the curve labeled with ``Proposed Small $t_{\mathsf{A}}$'' and ``Proposed Large $t_{\mathsf{A}}$'' in Fig.~\ref{fig:dec_er} give the DER of proposed method in the case of $t_{\mathsf{A}}=6/\alpha$ and the case of $t_{\mathsf{A}} = \infty$, respectively.
As shown in Fig.~\ref{fig:dec_er}, all the DERs are nearly equal.

As show in Remark \ref{rem:1}, $t_{\mathsf{A}}$ is a parameter for timeout, and in the case of $t_{\mathsf{A}} = 6/\alpha$, the decoding performance is slightly degraded compared with the existing one.

\subsection{Number of Decoding Processes \label{sec:hyo_con}}
In this section, we first evaluate the number of decoding processes and the number of updating edges at each iteration for the proposed algorithm.
Next, we compare the number of decoding processes at bit-wise decoding for each overhead.
\begin{figure}[tb]
  \centering
  \includegraphics[width=.785\linewidth]{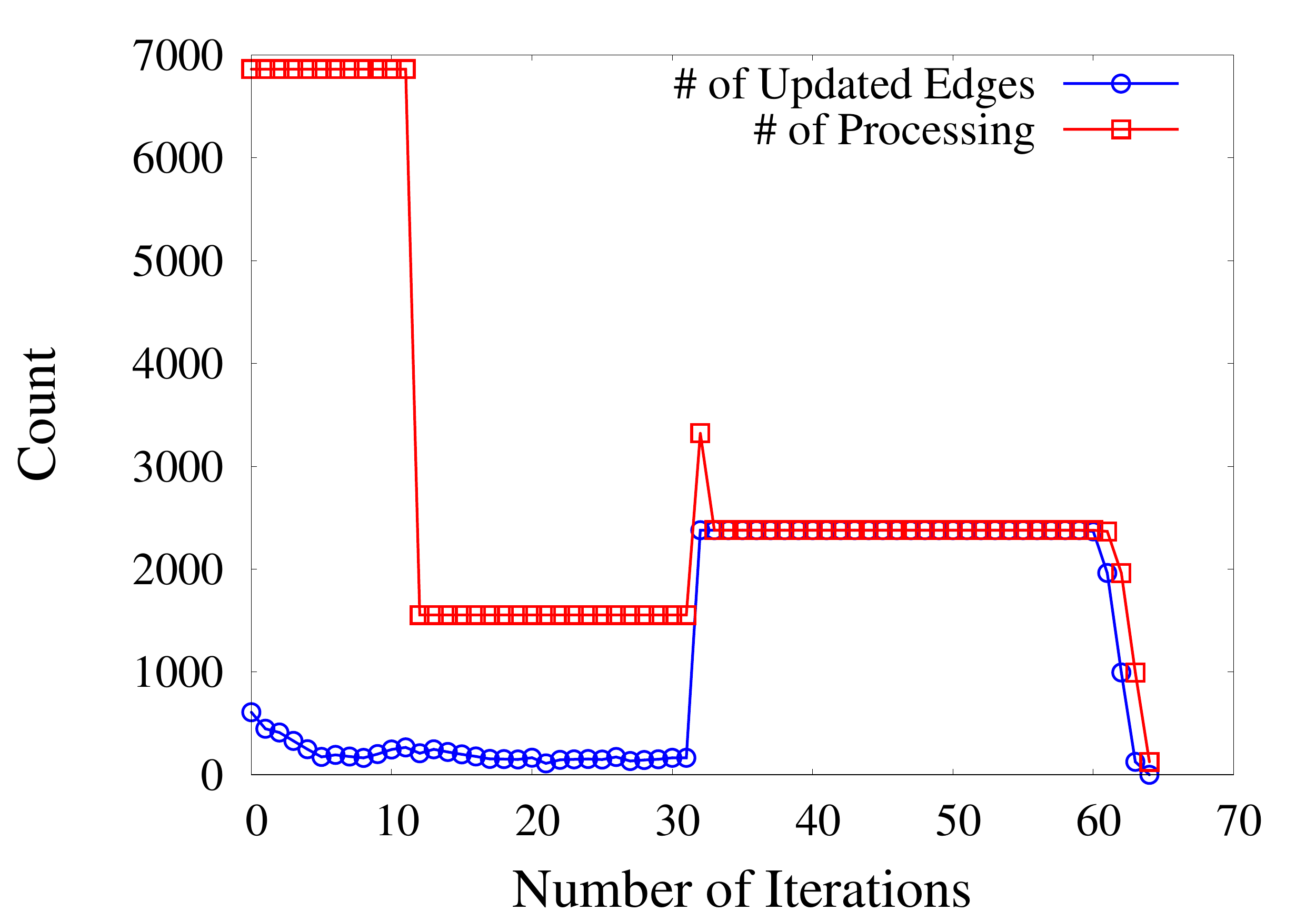}
  \caption{Number of decoding processes at each iteration for the proposed algorithm ($\alpha=0.10, \ell=100$) \label{fig:countL}}
\end{figure}

Figure \ref{fig:countL} displays an example of the number of decoding processes and number of updating edges at each iteration for the proposed algorithm.
The horizontal axis of Fig.~\ref{fig:countL} represents decoding iteration.
As shown in Fig.~\ref{fig:countL}, from the 12th iteration to the 31st iteration, the number of decoding processes at each iteration number is significantly reduced, because decoding processes are only executed the edges in $\mathcal{L}^{\mathsf{A}}$.
After 32nd iteration, the proposed algorithm executes the decoding process to the edges in $\mathcal{L}^{\mathsf{B}}$.
From Fig.~\ref{fig:countL}, we see that the number of updating edges almost equals to the number of decoding processes after 32nd iteration.
Hence, we conclude that the proposed algorithm well constructs edge list $\mathcal{L}^{\mathsf{B}}$.
Moreover, the number of iterations required for decoding is smaller than that of the existing algorithm.
Therefore, the proposed algorithm has fewer decoding processes than the existing algorithm.

\begin{figure}[tb]
  \centering
  \includegraphics[width=.785\linewidth]{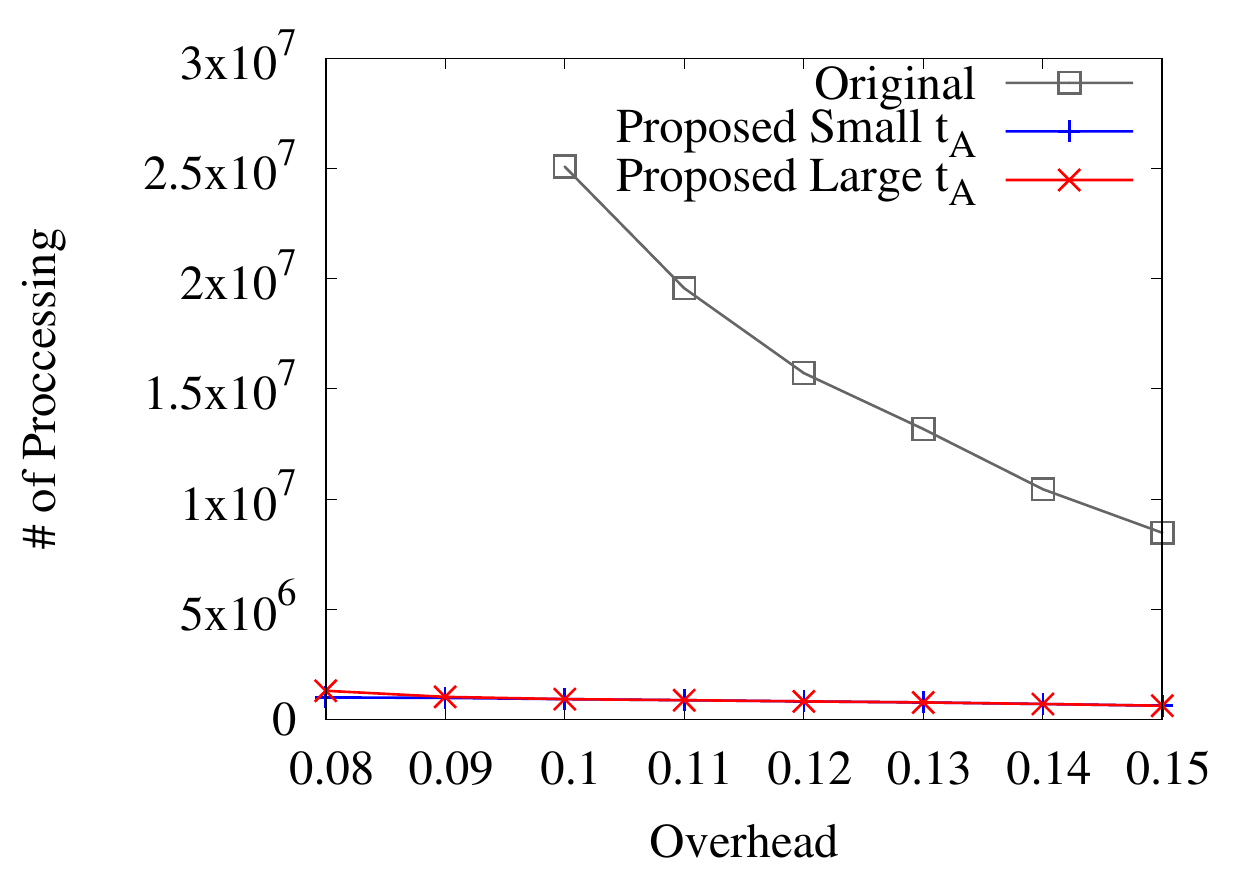}
  \caption{Comparison of the number of decoding processes for the proposed algorithm with existing one (Original) \label{fig:countT}}
\end{figure}

Next, we evaluate the number of decoding processes of an existing algorithm and the proposed algorithm for each overhead $\alpha$.
Figure \ref{fig:countT} compares the number of decoding processes for existing algorithm with proposed algorithm under $\ell=1000$.
The horizontal axis of Fig.~\ref{fig:countT} represents overhead $\alpha$.
From Fig.~\ref{fig:countT}, the number of decoding processes of the proposed algorithm is significantly less than one of an existing algorithm.

\subsection{Decoding Time\label{sec:hyo_time}}
\begin{figure}[t]
  \centering
   \includegraphics[width=.785\linewidth]{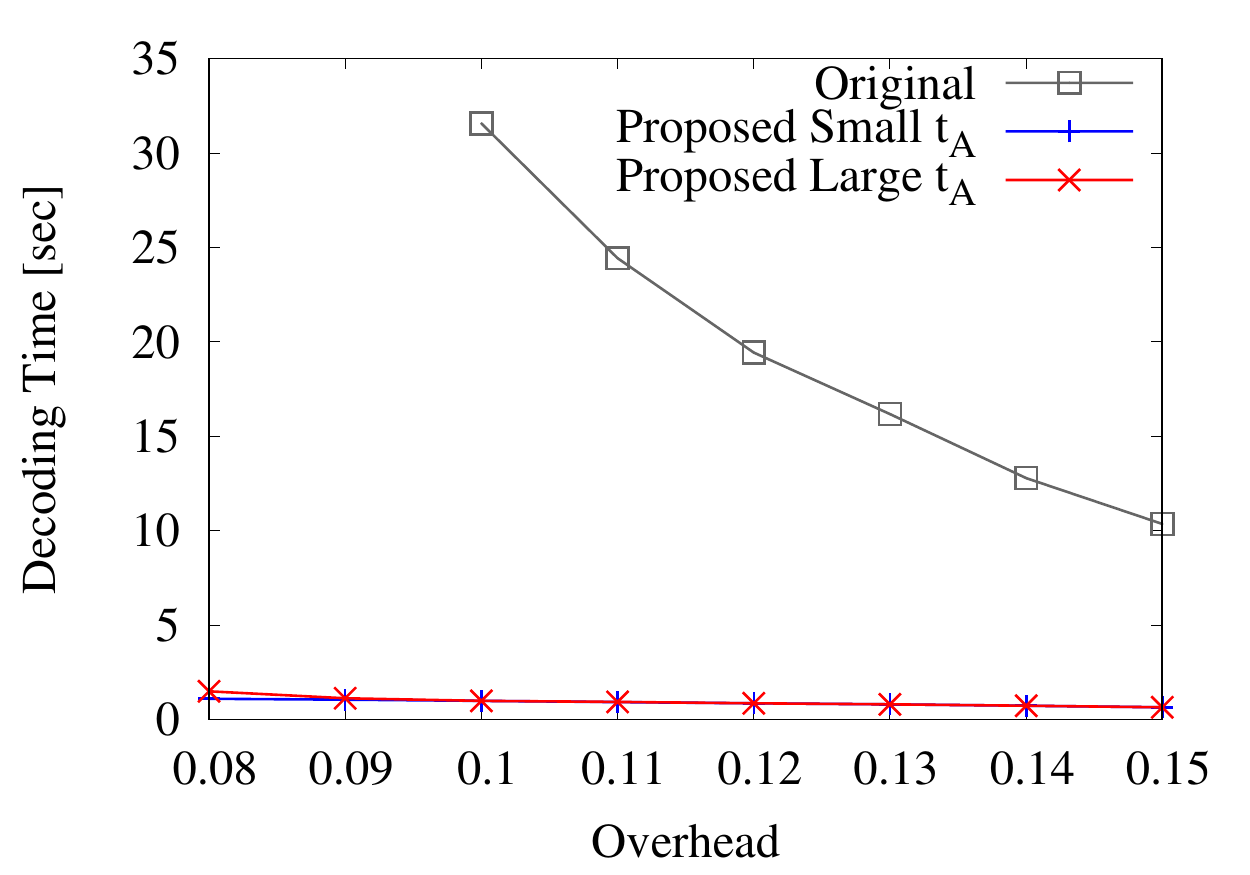}
   \caption{Comparison of the decoding time for the proposed algorithm with existing one (Original) \label{fig:Dec_time}}
\end{figure}

In the evaluation of this section, we try 10000 times for each overhead $\alpha$.
In this simulation, we use Ubuntu16.04 for OS, Intel(R)Core(TM)i7-4770CPU@3.40GHz for CPU, and 4GB DDR3 memory.
Figure \ref{fig:Dec_time} displays the decoding time for existing algorithm and proposed algorithm with $\ell=1000$.
The horizontal axis of Fig.~\ref{fig:Dec_time} represents overhead $\alpha$.
As shown in Fig.~\ref{fig:Dec_time}, the decoding time of the proposed algorithm is much shorter than one of existing one.
\section{Conclusion \label{sec:5}}
In this paper, we have proposed an efficient bit-wise decoding algorithm for ZDF codes.
Simulation results show that the proposed algorithm drastically reduces the decoding time compared with an existing algorithm.

\section*{Acknowledgment}
This work was supported by JSPS KAKENHI Grant Number 16K16007.

\end{document}